\documentclass[aps,pra,showpacs,twocolumn]{revtex4-1}
\usepackage{amssymb}
\usepackage{amsmath}
\usepackage{graphicx}
\usepackage{epsfig}
\usepackage{subfigure}
\usepackage{color}

\begin{document}

\title{Hidden exceptional point, localization-delocalization phase
transition in Hermitian bosonic Kitaev model}
\author{D. K. He}
\author{Z. Song}
\email{songtc@nankai.edu.cn}

\begin{abstract}
Exceptional points (EPs), a unique feature of non-Hermitian systems,
represent degeneracies in non-Hermitian operators that likely do not occur
in Hermitian systems. Nevertheless, unlike its fermionic counterpart, a
Hermitian bosonic Kitaev model supports a non-Hermitian core matrix,
involving a quantum phase transition (QPT) when an exceptional point
appears. In this study, we examine QPTs by mapping the Hamiltonian onto a
set of equivalent single-particle systems using a Bardeen-Cooper-Schrieffer
(BCS)-like pairing basis. We demonstrate the connection between the hidden
EP and the localization-delocalization transition in the equivalent systems.
The result is applicable to a Dicke model, which allows the experimental
detection of the transition based on the measurement of the average number
of photons for the quench dynamics stating from the empty state. Numerical
simulations of the time evolution reveal a clear transition point at the EP.
\end{abstract}

\affiliation{School of Physics, Nankai University, Tianjin 300071, China}
\maketitle

\section{Introduction}

{An exclusive feature of non-Hermitian systems is the concept of exceptional
points (EPs) \cite{kato1966,berry2004,heiss2012}, which are the degeneracies
of non-Hermitian operators. }The dynamics of the system with parameters far
away from, near and at the EP, exhibits extremely different behaviors \cite%
{yang2018dynamical,zhang2020resonant,zhang2019helical}. Consider a $2\times
2 $\ non-Hermitian matrix as an illustrative example. The matrix has two
identical eigenvectors at the EP. (i) When the system is far from or near an
EP and has a finite energy gap $\epsilon $, the dynamics exhibit periodic
oscillations, with the associated Dirac probability oscillating over a
period of $2\pi /\epsilon $. (ii) When the system is at an EP, the Dirac
probability increases quadratically with time. (iii) When the system has
complex energy levels, the Dirac probability increases exponentially with
time. The distinctive characteristics surrounding the EP have ignited
significant interest in both classical and quantum photonic systems \cite%
{Doppler2016,Xu2016,Assawaworrarit2017,Wiersig2014,Wiersig2016,Hodaei2017,Chen2017}%
.

A question that may seem naive but is actually insightful is whether there
is an EP hidden within a Hermitian system. {The previous works \cite%
{jin2010physics, jin2011partitioning,jin2011a,zhang2013self} for
non-interacting systems indicates that there exists a connection between the
EPs of an effective non-Hermitian system on a finite lattice and the dynamic
behavior in a finite-size Hermitian system. It is referred to as EP
dynamics, which stands for the exclusive dynamic behaviors and properties of
a non-Hermitian system at EP. This motivates us to reveal the connection
between the EP and the dynamics for a interacting many-bady system.}

Recently, the study of the bosonic Kitaev model has received much attention%
\cite%
{McDonald_PRX,wang2019non,flynn2020deconstructing,del2022non,wang2022quantum,bilitewski2023manipulating,ughrelidze2024interplay,slim2024optomechanical,busnaina2024quantum,hu2024bosonic}. In contrast to its fermionic counterpart, the Hermitian bosonic Kitaev
model features a non-Hermitian core matrix within its Nambu representation. 
It appears that the EP is not a unique feature of non-Hermitian systems. Although such a non-Hermitian matrix cannot result in a complex
quasi-particle spectrum of the original Hermitian Hamiltonian, special
dynamic behaviors may be associated with the emergence of its EP. It is our
goal to establish the connection between hidden EP and the possible phase
transition in dynamic behaviors of the system.

In this work, we examine the QPTs that arise from the EPs in a Hermitian
bosonic Kitaev model by exploring the exact solutions. It is demonstrated
that an EP exists in the core matrix with a given momentum, which results in
a non-analytic point in the solutions. We map the Hamiltonian onto a set of
equivalent single-particle systems using a Bardeen-Cooper-Schrieffer
(BCS)-like pairing basis. We demonstrate the connection between the hidden
EP and the localization-delocalization transition in the equivalent systems.
In addition, we apply the result to a Dicke model, which allows the
experimental detection of the transition based on the measurement of the
average photon number for the time evolution of the initial empty state.

Our results provide an alternative approach for bosonic Kitaev models. The
structure of this paper is as follows. In Sec. \ref{Model}, we introduce the
model and present the phase diagram in Sec. \ref{Phase diagram}. In Sec. \ref%
{Localization-delocalization transition}, we investigate the
localization-delocalization transition in the Fock space. In Sec. \ref%
{Quench dynamics in Dicke models}, we delve into the EP-associated
transition within the Dicke model, focusing on the quenching dynamics.
Finally, in Sec. \ref{Summary}, we provide a summary and discussion.

\section{Model}

\label{Model} The fermionic Kitaev chain is a one-dimensional tight-binding
model with the nearest-neighbor hopping and the superconducting $p$-wave
pairing terms on each dimer \cite{kitaev2001unpaired}. The corresponding
bosonic Kitaev model on $N$ lattice with imaginary hopping and pairing
amplitudes can be written as
\begin{equation}
H=\sum_{j=1}^{N}[itb_{j+1}^{\dag }b_{j}+i\Delta b_{j+1}^{\dag }b_{j}^{\dag }+%
\text{H.c.}+\mu \left( 2b_{j}^{\dag }b_{j}+1\right) ],  \label{H}
\end{equation}%
where $b_{j}^{\dag }$ $\left( b_{j}\right) $ is a bosonic creating
(annihilation) operator on site $j$. The periodic boundary condition is
imposed by defining $b_{1}=b_{N+1}$. Here, the hopping strength $t$ and
chemical potential $\mu $ are always real in this work. In this work, we
only consider the case with $\mu >0$\ and real $\Delta $\ for the sake of
simplicity. Using the Fourier transformation%
\begin{equation}
b_{j}=\frac{1}{\sqrt{N}}\sum_{-\pi \leqslant k<\pi }e^{ikj}b_{k}.
\end{equation}%
we get the Hamiltonian in $k$-space in the form%
\begin{eqnarray}
H &=&2\sum_{0\leqslant k<\pi }[\left( \mu +t\sin k\right) b_{k}^{\dag
}b_{k}+\left( \mu -t\sin k\right) b_{-k}b_{-k}^{\dag }  \notag \\
&&+i\Delta \cos kb_{k}^{\dag }b_{-k}^{\dag }+\text{H.c.}].
\end{eqnarray}%
In order to get the solution of the Hamiltonian, we rewrite it in the block
diagonal form $H=\sum_{0\leqslant k<\pi }H_{k}$, satisfying the commutation
relation $\left[ H_{k},H_{k^{\prime }}\right] =0$.\ It differs slightly from
the fermionic Kitaev model in that the Nambu representation of $H_{k}$ equals%
\begin{equation}
H_{k}=2\left(
\begin{array}{cc}
b_{k}^{\dag } & -b_{-k}%
\end{array}%
\right) h_{k}\left(
\begin{array}{c}
b_{k} \\
b_{-k}^{\dag }%
\end{array}%
\right) ,
\end{equation}%
where the core matrix $h_{k}$ is \cite{McDonald_PRX}%
\begin{equation}
h_{k}=\left(
\begin{array}{cc}
\mu & i\Delta _{k} \\
i\Delta _{k} & -\mu%
\end{array}%
\right) +T_{k},
\end{equation}%
with $T_{k}=t\sin k$ and $\Delta _{k}=\Delta \cos k$. We observed that the
matrix $h_{k}$ is non-Hermitian and becomes a Jordan block
\begin{equation}
h_{k}^{\mathrm{JD}}=\mu \left(
\begin{array}{cc}
1 & i \\
i & -1%
\end{array}%
\right) +T_{k},
\end{equation}%
satisfying $\left( h_{k}^{\mathrm{JD}}-T_{k}\right) ^{2}=0$, when taking $%
\mu =\Delta _{k}$, which is referred to as EP of the matrix. $h_{k}^{\mathrm{%
JD}}$ is not diagonalizable and possesses a coalescing eigenvector $\left(
-i,1\right) $. It is evident that the Hermiticity of $h_{k}$ within the
Hermitian Hamiltonian $H$ does not preclude the possibility of non-Hermitian
dynamics. Therefore, the existence of EP for the non-Hermitian matrix $h_{k}$%
\ should affect the property of the solutions, resulting a sudden change of
the corresponding dynamics of $H$. The main goal of this work is to explore
the connection between the EP of $h_{k}$ and the transition point of the
dynamics.

\section{Phase diagram}

\label{Phase diagram} Let us firstly determine the phase diagram by studying
the eigen problem of the matrix $h_{k}$. For the case with $\mu ^{2}\neq
\Delta _{k}^{2}$, taking a similarity transformation
\begin{equation}
V_{k}=\left(
\begin{array}{cc}
-i\sinh \frac{\theta _{k}}{2} & \cosh \frac{\theta _{k}}{2} \\
\cosh \frac{\theta _{k}}{2} & i\sinh \frac{\theta _{k}}{2}%
\end{array}%
\right) ,
\end{equation}%
the\ matrix $h_{k}$ can be diagonalized as%
\begin{equation}
\left( V^{k}\right) ^{-1}h_{k}V^{k}=\left(
\begin{array}{cc}
\varepsilon _{-}^{k} & 0 \\
0 & \varepsilon _{+}^{k}%
\end{array}%
\right) +T_{k},
\end{equation}%
with the eigenvalue%
\begin{equation}
\varepsilon _{\pm }^{k}=\pm 2\sqrt{\mu ^{2}-\Delta _{k}^{2}}.
\end{equation}%
Here the parameter $\theta _{k}$\ is determined by%
\begin{equation}
\tanh \frac{\theta _{k}}{2}=\frac{\mu -\varepsilon _{+}^{k}}{\Delta _{k}},
\end{equation}%
which can be complex. Based on the results, the Hamiltonian can be written
as the form%
\begin{equation}
H=2\sum_{-\pi \leqslant k<\pi }[\left( \varepsilon _{+}^{k}-T_{k}\right)
\left( \overline{\gamma }_{k}\gamma _{k}+\frac{1}{2}\right) ,
\label{H_gamma_k}
\end{equation}%
where the complex Bogoliubov modes defined as%
\begin{eqnarray}
\gamma _{k} &=&i\sinh \frac{\theta _{k}}{2}b_{k}^{\dag }+\cosh \frac{\theta
_{k}}{2}b_{-k},  \notag \\
\overline{\gamma }_{k} &=&-i\sinh \frac{\theta _{k}}{2}b_{k}+\cosh \frac{%
\theta _{k}}{2}b_{-k}^{\dag },
\end{eqnarray}%
and%
\begin{eqnarray}
\gamma _{-k} &=&i\sinh \frac{\theta _{k}}{2}b_{-k}^{\dag }+\cosh \frac{%
\theta _{k}}{2}b_{k},  \notag \\
\overline{\gamma }_{-k} &=&-i\sinh \frac{\theta _{k}}{2}b_{-k}+\cosh \frac{%
\theta _{k}}{2}b_{k}^{\dag },
\end{eqnarray}%
satisfying the canonical relations%
\begin{eqnarray}
\left[ \gamma _{k},\overline{\gamma }_{k^{\prime }}\right] &=&\delta
_{kk^{\prime }},  \notag \\
\left[ \gamma _{k},\gamma _{k^{\prime }}\right] &=&\left[ \overline{\gamma }%
_{k},\overline{\gamma }_{k^{\prime }}\right] =0,  \label{comm relation}
\end{eqnarray}%
with $k\in (-\pi ,\pi ]$.

It appears that the Hamiltonian $H$ in Eq. (\ref{H_gamma_k}) is in diagonal
form. However, an additional necessary condition for diagonalization is the
existence of a vacuum state $\left\vert \text{Vac}\right\rangle $ for the
operator $\gamma _{k}$, such that $\gamma _{k}\left\vert \text{Vac}%
\right\rangle =0$. When the vacuum state of the operator $\gamma _{k}$\ is
given, all the eigenstates can be constructed. In situations where the
explicit expression of the vacuum state is not available, one can
investigate the existence of such a state in the following manner. (i) In
the case where $\mu >\left\vert \Delta _{k}\right\vert $, direct derivation
shows that $\theta _{k}$\ is real for any $k$, which results in $\overline{%
\gamma }_{k}=\gamma _{k}^{\dag }$. Subsequently, the Hamiltonian $H$\ can be
expressed in the diagonal form%
\begin{equation}
H=2\sum_{-\pi \leqslant k<\pi }[\left( \varepsilon _{+}^{k}-T_{k}\right)
\left( \widehat{\mathcal{N}}_{k}+\frac{1}{2}\right) .
\end{equation}%
where $\widehat{\mathcal{N}}_{k}=\gamma _{k}^{\dag }\gamma _{k}$ is the
quasi-boson number operator. If the vacuum state $\left\vert \text{Vac}%
\right\rangle $\ exists, the energy levels are
\begin{equation}
E_{k}=2\left( \varepsilon _{+}^{k}-T_{k}\right) \left( \mathcal{N}_{k}+\frac{%
1}{2}\right) ,
\end{equation}%
where integers $\mathcal{N}_{k}\in \lbrack 0,\infty ]$, and $k\in (-\pi ,\pi
]$. Notably, the energy levels are equidistant with level spacing $%
\varepsilon _{+}^{k}+T_{k}$ for fixed $k$. These results are self-consistent
with the fact that the Hamiltonian $H$ is Hermitian. (ii) In contrast, in
the case where $\mu <\left\vert \Delta _{k}\right\vert $, we note that $%
\varepsilon _{+}^{k}$\ is imaginary.\ If the vacuum state $\left\vert \text{%
Vac}\right\rangle $\ exists, it is also the eigenstate of $H$, that is,%
\begin{equation}
H\left\vert \text{Vac}\right\rangle =i\sum_{-\pi \leqslant k<\pi }\left\vert
\varepsilon _{+}^{k}\right\vert \left\vert \text{Vac}\right\rangle .
\end{equation}%
Obviously, the presence of imaginary parts in the eigenvalues is
self-consistent with the fact that the Hamiltonian $H$ is Hermitian. It
implies that the vacuum state $\left\vert \text{Vac}\right\rangle $ does not
exist, and the corresponding Bogoliubov transformation is not the proper
method for diagonalization when $\mu <\left\vert \Delta _{k}\right\vert $.

This analysis predicts that the existence of EP for the core matrix $h_{k}$
will be reflected in the eigenstates and dynamics of the system. In the
following section, we will demonstrate and verify this point from an
alternative perspective.

\section{Localization-delocalization transition}

\label{Localization-delocalization transition} In this section, we
investigate the Hamiltonian $H$ in Eq. (\ref{H}) based on its equivalent
Hamiltonian, which describes the same physical system under a different
representation. We map the Hamiltonian onto a set of single-particle systems
using a Bardeen-Cooper-Schrieffer (BCS)-like pairing basis. We demonstrate
the connection between the EP and the localization-delocalization transition
in the equivalent systems.

There are many invariant subspaces for $H_{k}$ arising from the symmetries%
\begin{equation}
\left[ \Pi ,H_{k}\right] =\left[ K,H_{k}\right] =0,
\end{equation}%
where two operators are the boson number parity operator%
\begin{equation}
\Pi =(-1)^{n_{k}+n_{-k}},
\end{equation}%
and total momentum operator%
\begin{equation}
K=k\left( n_{k}-n_{-k}\right) ,
\end{equation}%
with $n_{k}=b_{k}^{\dag }b_{k}$. Here, we consider the case of a simply
invariant subspace, which is spanned by a basis set $\left\{ \left\vert
l\right\rangle ,l\in \lbrack 0,\infty )\right\} $%
\begin{equation}
\left\vert l\right\rangle =\frac{1}{l!}\left( b_{k}^{\dag }b_{-k}^{\dag
}\right) ^{l}\left\vert 0\right\rangle _{k}\left\vert 0\right\rangle _{-k}
\end{equation}%
with $\Pi \left\vert l\right\rangle =\left\vert l\right\rangle $\ and $%
K\left\vert l\right\rangle =0\left\vert l\right\rangle $, where$\ \left\vert
0\right\rangle _{k}\ $is the\ vacuum state for boson operator $b_{k}$, i.e.,
$b_{k}\left\vert 0\right\rangle _{k}=0$. Based on this set of basis, the
equivalent Hamiltonian of $H_{k}$\ can be expressed as the form%
\begin{equation}
H_{\mathrm{eq}}^{k}=i2\Delta _{k}\sum_{l=0}^{\infty }[(\left( l+1\right)
\left\vert l+1\right\rangle \left\langle l\right\vert -\text{H.c.})+2\mu
l\left\vert l\right\rangle \left\langle l\right\vert ],  \label{H_k_eq}
\end{equation}%
which describes a single-particle chain with an $l$-dependent
nearest-neighbor (NN) hopping term and linear potential. We still consider
the solutions of the equivalent Hamiltonian in the two regions, separated by
the EP.

\begin{figure*}[tbh]
\centering
\includegraphics[width=1\textwidth]{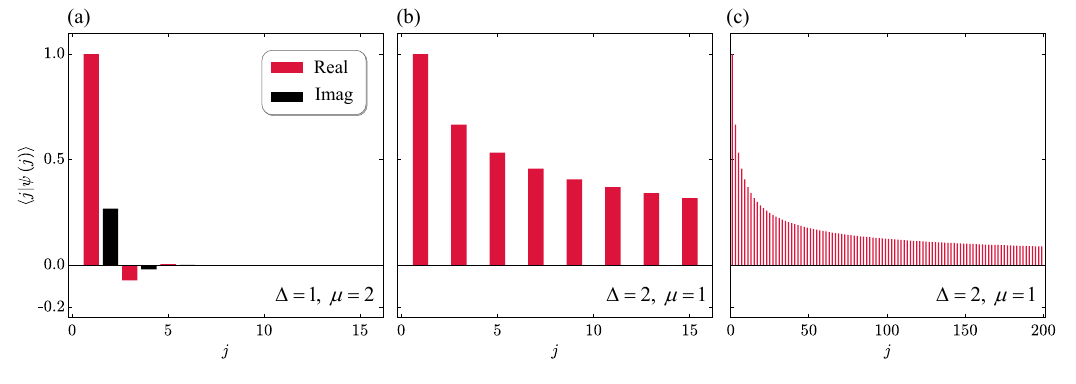}
\caption{Plots of two eigenstates $\left\vert \protect\psi \right\rangle $
of the eqivalent Hamiltonian $H_{\mathrm{eq}}^{k}$, given by Eq. (\protect
\ref{H_k_eq}), with two typical system parameters (a) $\Delta _{k}=1$ and $%
\protect\mu =2$, and (b, c) $\Delta _{k}=2$ and $\protect\mu =1$,
respectively. Here, the red or black bars represent the real or imaginary
parts of the amplitudes, respectively, and all are analytically determined.
(a) The profile of the vacuum state ($\left\vert \protect\psi \right\rangle
=\left\vert \mathrm{Vac}\right\rangle $) is obtained from Eq. (\protect\ref%
{Vac}), which is also the ground state with energy $E_{\mathrm{Vac}%
}^{k}=-0.5359$. It decays exponentially and then a localized state. (b, c)
is the profile of the eigenstate ($\left\vert \protect\psi \right\rangle
=\left\vert \protect\varphi _{k}\right\rangle $) with energy $E_{k}=0.0$,\
which is obtained from Eq. (\protect\ref{E=0 state}). (b) is the plot on the
same scale as (a) for comparison, while (b) is the same plot but on longer
scale. We can see form (c) that it becomes to a constant for large $j$,
indicating a delocalized state. These results imply that the EP at $\Delta
_{k}=\protect\mu $ is a transition point from localization to delocalization
in the Fock space.}
\label{fig1}
\end{figure*}

(i) In the case where $\mu >\left\vert \Delta _{k}\right\vert $, a
straightforward derivation shows that the state

\begin{equation}
\left\vert \mathrm{Vac}\right\rangle =\sum_{l=0}^{\infty }\left( i\right)
^{l+1}\left( -\frac{E_{\mathrm{Vac}}^{k}}{2\Delta _{k}}\right)
^{l}\left\vert l\right\rangle ,  \label{Vac}
\end{equation}%
is the eigenstate of $H_{\mathrm{eq}}^{k}$, satisfying $H_{\mathrm{eq}%
}^{k}\left\vert \mathrm{Vac}\right\rangle =E_{\mathrm{Vac}}^{k}\left\vert
\mathrm{Vac}\right\rangle $ with energy
\begin{equation}
E_{\mathrm{Vac}}^{k}=2\sqrt{\mu ^{2}-\Delta _{k}^{2}}-2\mu ,
\end{equation}%
under the condition $\mu ^{2}\geqslant \Delta _{k}^{2}$. Importantly, it is
just the vacuum state of the operator $\gamma _{k}$, i.e.,
\begin{equation}
\gamma _{k}\left\vert \mathrm{Vac}\right\rangle =0,
\end{equation}%
based on which, other eigenstats of $H_{\mathrm{eq}}^{k}$\ can be
constructed as $\left( \gamma _{k}^{\dag }\gamma _{-k}^{\dag }\right)
^{n}\left\vert \mathrm{Vac}\right\rangle $ for integer $n$, with energy
\begin{equation}
E_{n}^{k}=4n\sqrt{\mu ^{2}-\Delta _{k}^{2}}-2\mu .
\end{equation}%
The profile of the wave function of $\left\vert \mathrm{Vac}\right\rangle $
is plotted in Fig. \ref{fig1}(a). It indicates that state $\left\vert
\mathrm{Vac}\right\rangle $ is a localized state.

(ii) In the case where $\mu <\left\vert \Delta _{k}\right\vert $, the above
solution is invalid, and we must then seek another way. To proceed, we
introduce another boson operator%
\begin{equation}
\bar{b}_{k}=i\kappa _{+}b_{k}^{\dag }+\kappa _{-}b_{-k},
\end{equation}%
with%
\begin{equation}
\kappa _{\pm }=\frac{\sqrt{\left\vert \Delta _{k}\right\vert -\mu }\mp \sqrt{%
\left\vert \Delta _{k}\right\vert +\mu }}{2\left( \Delta _{k}^{2}-\mu
^{2}\right) ^{1/4}},
\end{equation}%
to rewrite $H^{k}$\ in the form

\begin{eqnarray}
H_{k} &=&i\sqrt{\Delta _{k}^{2}-\mu ^{2}}\left( \bar{b}_{-k}^{\dag }\bar{b}%
_{k}^{\dag }-\bar{b}_{k}\bar{b}_{-k}\right)  \notag \\
&&+T_{k}\left( \bar{b}_{-k}^{\dag }\bar{b}_{-k}-\bar{b}_{k}^{\dag }\bar{b}%
_{k}\right) -\mu .
\end{eqnarray}%
Similarly, we consider the case of a simply invariant subspace, which is
spanned by a basis set $\left\{ \overline{\left\vert l\right\rangle },l\in
\lbrack 0,\infty )\right\} $%
\begin{equation}
\overline{\left\vert l\right\rangle }=\frac{1}{l!}\left( \bar{b}_{k}^{\dag }%
\bar{b}_{-k}^{\dag }\right) ^{l}\overline{\left\vert 0\right\rangle }_{k}%
\overline{\left\vert 0\right\rangle }_{-k},
\end{equation}%
where$\ \left\vert 0\right\rangle _{k}\ $is the\ vacuum state for boson
operator $\bar{b}_{k}$, i.e., $\bar{b}_{k}\overline{\left\vert
0\right\rangle }_{k}=0$. Based on this set of basis, the equivalent
Hamiltonian of $H_{k}$\ can be expressed as the form%
\begin{eqnarray}
\overline{H}_{\mathrm{eq}}^{k} &=&i2\sqrt{\Delta _{k}^{2}-\mu ^{2}}%
\sum_{l=0}^{\infty }(\left( l+1\right) \overline{\left\vert l+1\right\rangle
}\overline{\left\langle l\right\vert }  \notag \\
&&-\text{H.c.})-\mu ,
\end{eqnarray}%
which describes a single-particle chain with an $l$-dependent
nearest-neighbor (NN) hopping term and zero linear potential. The
corresponding Schrodinger equation is
\begin{equation}
\overline{H}_{\mathrm{eq}}^{k}\left\vert \varphi _{k}\right\rangle
=E_{k}\left\vert \varphi _{k}\right\rangle .
\end{equation}%
The solution to such a semi-infinite chain can be exactly obtained as
follows. For any given real number $E_{k}$\ as the eigenvalues, the
corresponding eigenvector%
\begin{equation}
\left\vert \varphi _{k}\right\rangle =\sum_{l=0}c_{l}\overline{\left\vert
l\right\rangle },
\end{equation}%
is given by the following recursive formula%
\begin{eqnarray}
c_{0} &=&1,  \notag \\
c_{1} &=&i\frac{E_{k}}{\sqrt{\Delta _{k}^{2}-\mu ^{2}}},  \notag \\
&&...,  \notag \\
c_{l+1} &=&\frac{l\sqrt{\Delta _{k}^{2}-\mu ^{2}}c_{l-1}+iE_{k}c_{l}}{\left(
l+1\right) \sqrt{\Delta _{k}^{2}-\mu ^{2}}}  \notag \\
&&....
\end{eqnarray}%
Taking $E_{k}=0$,\ as an example, we have
\begin{eqnarray}
c_{0} &=&1,c_{1}=0,  \notag \\
c_{2} &=&\frac{1}{2},c_{3}=0,  \notag \\
&&...,  \notag \\
c_{2n} &=&\frac{\left( 2n-1\right) }{2n}c_{2\left( n-1\right) },c_{2n-1}=0
\notag \\
&&...,  \label{E=0 state}
\end{eqnarray}%
which is plotted in Fig. \ref{fig1}(b). The profile of the wave function
indicates that the corresponding eigen state is an extended state in the
space spanned by the basis set $\left\{ \overline{\left\vert l\right\rangle }%
\right\} $.\ Although there is no rigorous proof, the following relation%
\begin{eqnarray}
&&\left\langle l\right\vert (\bar{b}_{k}^{\dag }\bar{b}_{-k}^{\dag }-\bar{b}%
_{k}\bar{b}_{-k})\left\vert l^{\prime }\right\rangle  \notag \\
&=&\left( \kappa _{+}^{2}+\kappa _{-}^{2}\right) \left( l\delta
_{l,l^{\prime }-1}-l^{\prime }\delta _{l,l^{\prime }+1}\right) -i4\kappa
_{-}\kappa _{+}l\delta _{l^{\prime },l},
\end{eqnarray}%
implies that an extended state in the space $\left\{ \overline{\left\vert
l\right\rangle }\right\} $\ is also an extended state in the space $\left\{
\left\vert l\right\rangle \right\} $.

\begin{figure}[tbph]
\centering
\includegraphics[width=0.45 \textwidth]{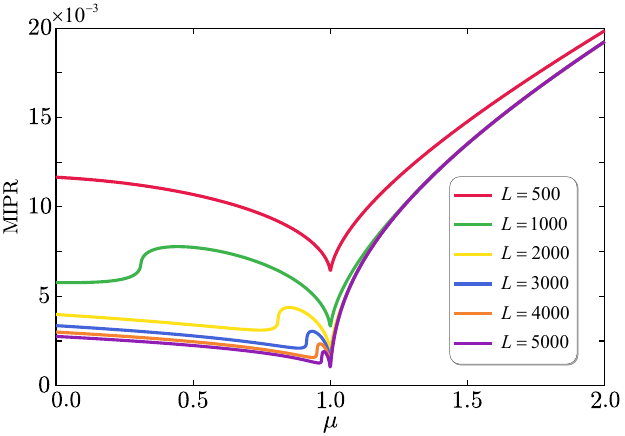}
\caption{Plot of \textrm{MIPR}, which represents the arithmetic mean of the
\textrm{IPRs} for a set of eigenstates of the Hamiltonian $H_{\mathrm{eq}%
}^{k}$, given by Eq. (\protect\ref{H_k_eq}). The system parameters are $%
\Delta =1$ and $N$ indicated in the panel. The selected set of states
consists of the lowest $400$ eigenstates. It is evident that there is a
transition point at $\protect\mu =1$\ for every given $N$. Furthemore, the
value of $\mathrm{MIPR}$ becomes stable as $N$ increases.\ For $\protect\mu %
<1$, the asymptotic value of $\mathrm{MIPR}$ approaches zero as the system
size becomes sufficiently large, whereas it remains finite for $\protect\mu %
>1$.}
\label{fig2}
\end{figure}

The above analysis implies that the eigenstates of $H_{\mathrm{eq}}^{k}$ are
localized in one region ($\mu >\left\vert \Delta _{k}\right\vert $) and
extended in the other ($\mu <\left\vert \Delta _{k}\right\vert $). A natural
question arises: whether it is a universal conclusion for other eigenstates.
At this stage, we can only answer this question with the aid of numerical
simulations. To characterize the localization features, we use the inverse
participation ratio (\textrm{IPR}) of a given state $\left\vert \psi
\right\rangle $ as a criterion to distinguish the extended states from the
localized ones. The IPR is defined as%
\begin{equation}
\mathrm{IPR}=\frac{\sum_{l}\left\vert \left\langle \psi \right\vert l\rangle
\right\vert ^{4}}{(\sum_{l}\left\vert \left\langle \psi \right\vert l\rangle
\right\vert ^{2})^{2}}.
\end{equation}%
For spatially extended states, the value of $\mathrm{IPR}$ approaches zero
as the system size becomes sufficiently large, whereas it remains finite for
localized states, regardless of the system size. For example, the $\mathrm{%
IPR}$\ for the state $\left\vert \mathrm{Vac}\right\rangle $ given in Eq. (%
\ref{Vac}) can be estimated as\
\begin{equation}
\mathrm{IPR}_{\mathrm{V}}=\frac{\Delta _{k}^{2}}{\mu ^{2}-\mu \sqrt{\mu
^{2}-\left( \Delta _{k}\right) ^{2}}}-1,
\end{equation}%
which is finite, identifying it as a local state. We perform numerical
simulation to explore localization-delocalization transition for a trancted
chain with length $L$. We compute the mean inverse participation ratio
(MIPR), which is the arithmetic mean of the IPRs for all the eigenstates.
The finite size effect can be excluded in our analysis because the system
size is large enough to approximate an infinite system. The MIPR is plotted
as a function of the system parameter in Fig. \ref{fig2}, showing an evident
sudden change at the EP.

We would like to point out that the localization-delocalization transition
occurs in the Fock space and in some invariant subspaces. This is slightly
different from the usual transition in real space for systems such as the AA
model and the Anderson model, where the transition is entirely determined by
the system parameters.

\section{Quench dynamics in Dicke model}

\label{Quench dynamics in Dicke models}

\begin{figure*}[tbph]
\centering
\includegraphics[width=1\textwidth]{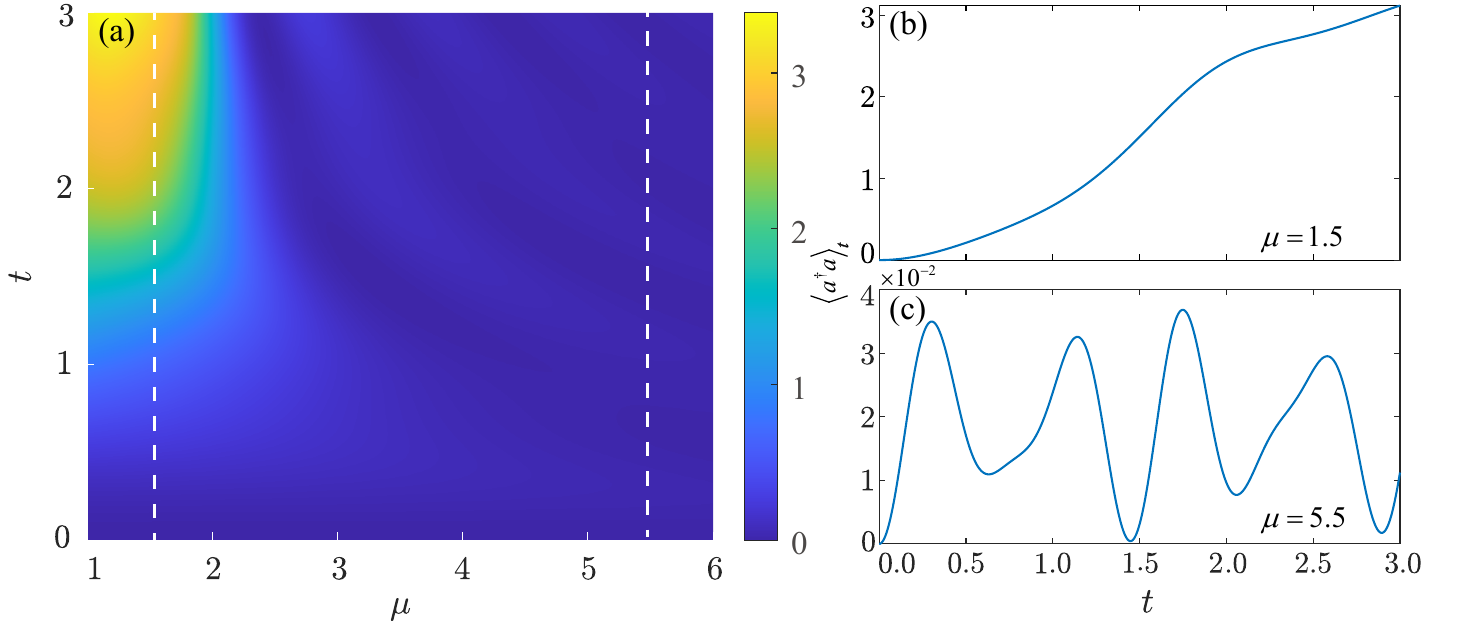}
\caption{Plots of the average photon number $N_{\text{\textrm{P}}}\left(
t\right) $, given by Eq. (\protect\ref{Np(t)}),\ for the quench process in
the Dicke model with the Hamiltonian $H_{\text{\textrm{D}}}$, as defined by
the Eq. (\protect\ref{H_D}). The numerical simulation is performed by
tracking the time evolution of the initial state $\left\vert \Psi \left(
0\right) \right\rangle $, which is detailed in the main text preceding Eq. (%
\protect\ref{Np(t)}). The quanity $N_{\text{\textrm{P}}}\left( t\right) $\
for a given $\protect\mu $\ is obtained by exact diagonalization of the
finite-dimensional matrix representation of $H_{\text{\textrm{D}}}$, under
the truncation of the photon number. The system parameters are an atom
number $N_{\text{\textrm{atom}}}=64$\ and $\Delta =1$. (a) A 3D plot of $N_{%
\text{\textrm{P}}}$\ as a function of time $t$ and $\protect\mu $.\ (b, c)
The profiles of $N_{\text{\textrm{P}}}\left( t\right) $\ for two typical
values of $\protect\mu $.}
\label{fig3}
\end{figure*}

In this section, we utilize the results derived in the previous sections to
construct a scheme for experimentally demonstrating the dynamic signature of
the EP in the Dicke model. It is a fundamental model of quantum optics,
which describes the interaction between light and matter. In the Dicke
model, the light component is described as a single quantum mode, while the
matter is described as a set of two-level systems. The model is
characterized by the following Hamiltonian for $N$ atoms%
\begin{equation}
H_{\text{\textrm{D}}}=\mu \left( a^{\dag }a+J_{z}\right) +\frac{\Delta }{%
\sqrt{N_{\text{\textrm{atom}}}}}\left( a^{\dag }+a\right) \left(
J_{+}+J_{-}\right) ,  \label{H_D}
\end{equation}%
where $a^{\dag }$ $\left( a\right) $ is a bosonic creating (annihilation)
operator. Here $\mu >0$\ is the cavity frequency, or the chemical potential
of the bosons and is also the energy difference between the states of each
two-level system, $\Delta >0$ characterizes the coupling between the
two-level systems and the cavity. Here, we use some special letters ($\mu
,\Delta $) for the sake of consistency across contexts. We define the
macroscopic spin operators $J_{\alpha }$ ($\alpha =x,y,z$) and $J_{\pm
}=J_{x}\pm iJ_{y}$\ of spin-$N/2$ to represent the collective state of $N$
atoms, which satisfy the spin algebra, $\left[ J_{\alpha },J_{\beta }\right]
=i\varepsilon _{\alpha \beta \gamma }J_{\gamma }$. Introducing the
Holstein-Primakoff boson representation of the collective spin operators%
\begin{eqnarray}
J_{z} &=&b^{\dag }b-\frac{N_{\text{\textrm{atom}}}}{2},  \notag \\
J_{+} &=&\left( J_{-}\right) ^{\dag }=b^{\dag }\sqrt{N_{\text{\textrm{atom}}}-b^{\dag }b},
\end{eqnarray}%
the Hamiltonian can be rewritten as the form%
\begin{equation}
H_{\text{\textrm{eff}}}=\mu \left( a^{\dag }a+b^{\dag }b\right) +\Delta
\left( a^{\dag }+a\right) \left( b^{\dag }+b\right) ,  \label{Hab}
\end{equation}%
by neglecting a constant term. It is exactly the same as the two-site
bosonic Kitaev model we studied in previous sections. Taking the transform%
\begin{equation}
d_{\pm }=\frac{1}{\sqrt{2}}\left( a\pm b\right) ,
\end{equation}%
the Hamiltonian can be expressed as
\begin{equation}
H_{\text{\textrm{eff}}}=\varphi _{\mathrm{L}}\left(
\begin{array}{cc}
h_{\text{\textrm{eff}}}^{+} & 0 \\
0 & h_{\text{\textrm{eff}}}^{-}%
\end{array}%
\right) \varphi _{\mathrm{R}},
\end{equation}%
with the non-Hermitian matrices%
\begin{eqnarray}
h_{\text{\textrm{eff}}}^{\pm } &=&\frac{\mu }{2}\left(
\begin{array}{cc}
1 & 0 \\
0 & -1%
\end{array}%
\right) \pm \frac{\Delta }{2}\left(
\begin{array}{cc}
1 & 1 \\
-1 & -1%
\end{array}%
\right)  \\
&=&\frac{\left( \mu \pm \Delta \right) }{2}\sigma _{z}\pm \frac{i\Delta }{2}%
\sigma _{y},
\end{eqnarray}%
where $\sigma _{z}$ and $\sigma _{y}$ are Pauli matrices, and the two
operator vectors are $\varphi _{\mathrm{L}}=\left( d_{+},-d_{+}^{\dagger
},d_{-},-d_{-}^{\dagger }\right) $ and $\varphi _{\mathrm{R}}=\left(
d_{+}^{\dagger },d_{+},d_{-}^{\dagger },d_{-}\right) ^{T}$. Obviously, the
EP occurs at $\mu =\mu _{c}=\Delta $, at which one of the two matrices $h_{%
\text{\textrm{D}}}^{+}$\ and $h_{\text{\textrm{D}}}^{-}$\ is undiagonable.
It should result in the critical behavior both in the ground state and
dynamics of $H_{\text{\textrm{D}}}$. In the case with $\mu \neq \Delta $,
the Hamiltonian can be diagonalized as the form%
\begin{equation}
H_{\text{\textrm{eff}}}=\sum_{\rho =\pm }\varepsilon _{\rho }\left(
\overline{\gamma }_{\rho }\gamma _{\rho }+\frac{1}{2}\right) ,\label{Dicke_H}
\end{equation}%
where the energy is%
\begin{equation}
\varepsilon _{\sigma \rho }=\frac{\sigma }{2}\sqrt{\mu ^{2}+2\rho \Delta \mu
},
\end{equation}%
and the complex Bogoliubov modes defined as%
\begin{eqnarray}
\gamma _{\rho } &=&\sinh \frac{\theta _{\rho }}{2}d_{\rho }^{\dag }+\cosh
\frac{\theta _{\rho }}{2}d_{\rho },  \notag \\
\overline{\gamma }_{\rho } &=&\sinh \frac{\theta _{\rho }}{2}d_{\rho }+\cosh
\frac{\theta _{\rho }}{2}d_{\rho }^{\dag },
\end{eqnarray}%
where\textbf{\ }$\tanh \left( \theta _{\rho }/2\right) =1$ $-\rho (\mu
/\Delta )(1+\sqrt{1+2\rho \Delta /\mu })$, satisfying the canonical
relations%
\begin{eqnarray}
\left[ \gamma _{\rho },\overline{\gamma }_{\rho ^{\prime }}\right]
&=&\delta _{\rho \rho ^{\prime }},  \notag \\
\left[ \gamma _{\rho },\gamma _{^{\prime }\rho ^{\prime }}\right]  &=&\left[
\overline{\gamma }_{\rho },\overline{\gamma }_{\rho ^{\prime }}\right] =0.
\end{eqnarray}%
As concluded before, such complex Bogoliubov modes are only available in the
region $\mu >\Delta $, in which we still have $\overline{\gamma }_{\pm
}=\gamma _{\pm }^{\dag }$. We investigate the solutions in the two regions,
respectively.

\begin{figure}[tbph]
	\centering
	\includegraphics[width=0.4\textwidth]{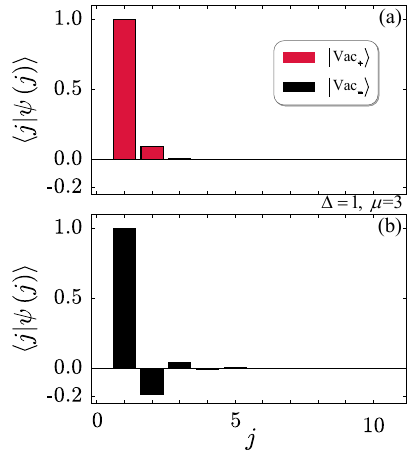}
	\caption{Plots of two eigenstates of the effective model Eq. (\ref{Hab}), specifically the vacuum state, as given by Eq. (\ref{Dicke_vac}), with the system parameters $\Delta=1$ and $\mu=3$ for figures (a) and (b). In these plots, the red or black bars represent the real or imaginary parts of the amplitudes, respectively, and all are analytically determined.}
	\label{fig4}
\end{figure}

(i) In the region $\mu >\Delta $,\ the vacuum state of the quasi-boson
operator $\gamma _{\pm }$\ is%
\begin{equation}
\left\vert \mathrm{Vac}_{\pm }\right\rangle =\sum_{l=0}^{\infty }\left( \eta
_{\pm }\right) ^{l}A_{l}\left\vert 2l\right\rangle \label{Dicke_vac},
\end{equation}%
with $\left\vert l\right\rangle =\frac{1}{\sqrt{2l}}\left( d_{\rho }^{\dag
}\right) ^{2l}\left\vert 0\right\rangle $, satisfying $\gamma _{\pm
}\left\vert \mathrm{Vac}_{\pm }\right\rangle =0$, where the coefficients are
given by%
\begin{equation}
\eta _{\pm }=\frac{\Delta \pm \mu \mp \sqrt{\mu ^{2}\pm 2\Delta \mu }}{\sqrt{%
2}\Delta },
\end{equation}%
and%
\begin{eqnarray}
A_{0} &=&A_{1}=1,  \notag \\
A_{l+2} &=&\sqrt{\frac{2l^{2}+5l+3}{2l^{2}+5l+2}}\frac{\left( A_{l+1}\right)
^{2}}{A_{l}}.
\end{eqnarray}%
The profiles of the vacuum states
\begin{equation}
p_{\pm }(l)=\frac{\left\vert \langle 2l\left\vert \mathrm{Vac}_{\pm
}\right\rangle \right\vert ^{2}}{\sum_{l=0}\left\vert \langle 2l\left\vert
\mathrm{Vac}_{\pm }\right\rangle \right\vert ^{2}},
\end{equation}%
for several represntative values of $\mu /\Delta $\ are plotted in Fig. \ref%
{fig4}. The corresponding \textrm{IPRs} indicate that $\left\vert \mathrm{%
Vac}_{\pm }\right\rangle $\ are localized state in the $\left\{ \left\vert
2l\right\rangle \right\} $\ space. It is easy to check that the ground state
of $H_{\text{\textrm{D}}}$\ within the region $\mu >\Delta $\ is the vacuum
state of two types of quasi-boson,

\begin{equation}
\left\vert \mathrm{G}\right\rangle =\left\vert \mathrm{Vac}_{+}\right\rangle
\left\vert \mathrm{Vac}_{-}\right\rangle ,
\end{equation}%
with the groundstate energy $\varepsilon _{+}+\varepsilon _{-}-\mu $,
because the energy $\varepsilon _{\pm }$ always positive.

(ii) In the region $\mu <\Delta $, we introduce another two independent
boson operators%
\begin{equation*}
A_{\rho }=\sinh \frac{\varphi _{\rho }}{2}d_{\rho }^{\dag }+\cosh \frac{%
\varphi _{\rho }}{2}d_{\rho }
\end{equation*}%
where $\tanh \left( \varphi _{\rho }/2\right) =\left( \rho \Delta +\sqrt{%
-\left( \mu ^{2}+2\rho \Delta \mu \right) }\right) /\left( \rho \Delta +\mu
\right) $, to rewrite $H_{\text{\textrm{eff}}}$\ in the form%
\begin{equation}
H_{\text{\textrm{eff}}}=\sum_{\rho =\pm }-\frac{1}{2}\sqrt{-\left( \mu
^{2}+2\rho \Delta \mu \right) }\left( A_{\rho }^{\dag }A_{\rho }^{\dag
}+A_{\rho }A_{\rho }\right) ,
\end{equation}%
which are exactly solvable based on the analysis in the last section.

A natural question arises: How can we detect the existence of the hidden EP
in the Dicke model? Specifically, how can we identify the transition point
between localized and extended phases experimentally? We begin with a
dynamical method to differentiate between extended and localized regions. As
shown in the last section, some localized eigenstates have a finite
probability on the vacuum state $\left\vert 0\right\rangle $ of the operator
$d_{\rho }$. In contrast, nonlocalized eigenstates have an infinitesimal
probability on the vacuum state of the operator $d_{\rho }$. This feature
forms the foundation for the dynamical detection of the transition.

\begin{figure}[tbph]
\centering
\includegraphics[width=0.5\textwidth]{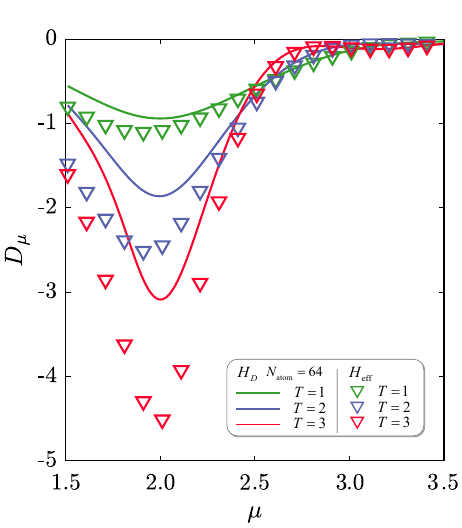}
\caption{Plots of $D_{\protect\mu }$, the derivative of the average photon
number over a period of time $T$ with respect to $\protect\mu $ ($\Delta =1$%
), as defined by Eq. (\protect\ref{Dm}).\ The numerical results are obtained
through exact diagonalizations of the matrix representations for both the
Dicke Hamitonian Eq. (\protect\ref{H_D}) and the effective Hamiltonian
Eq. (\protect\ref{Hab}), under the truncation of the photon number. The
time internval $T$\ and the number of atoms $N_{\text{\textrm{atom}}}$\ are
indicated in the panel. The results show that each $D_{\protect\mu }$\ has a
minimum near the EP at $\protect\mu =2$.\ The valleys become deeper as $T$
or $N_{\text{\textrm{atom}}}$\ increases, implying that $D_{\protect\mu }$\
is divergent at the EP\ in the case of infinite $T$ and $N_{\text{\textrm{%
atom}}}$.}
\label{fig5}
\end{figure}

We consider the time evolution of an initial state $\left\vert
0\right\rangle $. When $\mu > 2 \Delta $, it represents a superposition of a
set of localized eigenstates, while for $\mu < 2 \Delta $, it represents
nonlocalized eigenstates. In the former case, the profile of the evolved
state remains stationary or does not spread. Conversely, in the latter case,
the evolved state will move away or spread out. These differences should
result in two distinct dynamic behaviors for the average number of photons
in the Dicke model: it is finite when $\mu > 2 \Delta $, while it explodes when
$\mu < 2 \Delta $.

The experimental scheme involves considering a quench process from the empty
state $\left\vert 0\right\rangle _{\text{p}}\left\vert \downarrow
\right\rangle $, in which the photon number is zero and all $N$ atoms are in
the ground state, that is, $a\left\vert 0\right\rangle _{\text{p}}=0$\ and\ $%
J_{z}\left\vert \downarrow \right\rangle =-N/2$. The numerical simulation is
performed by computing the time evolution of an initial state $\left\vert
\Psi \left( 0\right) \right\rangle =\left\vert 0\right\rangle _{\text{p}%
}\left\vert \downarrow \right\rangle $ under the original Hamiltonian $H_{%
\mathrm{D}}$ given by the Eq. (\ref{H_D}). We use the average number of
photons for the evolved state $\left\vert \Psi \left( t\right) \right\rangle
$, which is given by
\begin{equation}
N_{\text{P}}(t)=\left\langle \Psi \left( t\right) \right\vert a^{\dag
}a\left\vert \Psi \left( t\right) \right\rangle =\left\vert ae^{-iH_{\mathrm{%
D}}t}\left\vert 0\right\rangle _{\text{p}}\left\vert \downarrow
\right\rangle \right\vert ^{2},  \label{Np(t)}
\end{equation}%
to measure the difference of the two phases separated by the EP. In order to
detect the accurate transition point, we calulate the derivative of the the
average photon number over a period of time $T$ with respect to $\mu $,

\begin{equation}
D_{\mu }=\frac{\partial }{\partial \mu }\left[ \frac{1}{T}\int_{0}^{T}N_{%
\text{P}}(t)\text{\textrm{d}}t\right] .  \label{Dm}
\end{equation}%
According to the analysis above, the average photon number should increase
as $T$ increases for $\mu <2\Delta $, while it should remain a finite
constant for $\mu >2\Delta $. Then, at the EP, the average photon number
should experience a jump, resulting in the divergence of $D_{\mu }$\ in the
limit as $T\rightarrow \infty $. It is expected that $D_{\mu }$\ has an
extremum near the EP. The numerical results presented in Fig. \ref{fig5}
demonstrate a good match between the dynamic behavior and the transition
point, indicating that our method is highly efficient.

\section{Summary}

\label{Summary}

In summary, we have revealed the possible connection between the QPT in a
Hermitian system and the EP hidden within it, based on the solutions of the
bosonic Kitaev model. As a candidate model to demonstrate the issue, such a
model has two properties: (i) The degree of freedom in the bosonic Kitaev
model is infinite, even for finite size system; (ii) The core matrix for its
Nambu representation is non-Hermitian. In this sense, the EPs is no longer a
unique feature for non-Hermitian systems. Technically, we demonstrate the
connection between the hidden EPs and the localization-delocalization
transition in equivalent systems by mapping the Hamiltonian onto a set of
equivalent single-particle systems using a BCS-like pairing basis. We also
propose a scheme for the experimental detection of the transition in a Dicke
model, based on the measurement of the average number of photons for the
quench dynamics starting from the empty state. Our work may open a new
research area for uncovering the hidden EPs in a Hermitian system that
govern the QPTs.

\section*{Acknowledgment}

We acknowledge the support of NSFC (Grants No. 12374461).

\end{document}